# A Noise Mitigation Approach for VLC Systems

Antonio Costanzo
Inria Lille – Nord Europe
Villeneuve d'Ascq, France
antonio.costanzo@inria.fr

Valeria Loscri
Inria Lille – Nord Europe
Villeneuve d'Ascq, France
*IEEE Senior Member*
valeria.loscri@inria.fr

Mauro Biagi
Sapienza University
Rome, Italy
*IEEE Senior Member*
mauro.biagi@uniroma1.it

*Abstract*—Visible Light Communication (VLC) is based on the dual use of the illumination infrastructure for wireless data communication. The major interest on this communication technology lies on its specific features to be a secure, cost-effective wireless technology. Recently, this technology has gained an important role as potential candidate for complementing traditional RF communication systems. Anyway a major issue for the VLC development is a deep comprehension of the noise and its impact on the received signal at the receiver. In this work, we present a simple but effective approach to analyze the noise and drastically reduce it through a signal processing method. In order to validate the effectiveness of this analytical approach, we have developed an USRP-based testbed. Experimental results have been carried out by evaluating the symbol error rate (SER) and show the effectiveness of the noise mitigation approach in different interference conditions and at different distance between the transmitter and the receiver.

## I. Introduction (Heading 1)

As one of the emerging optical wireless communication techniques, the visible light communication (VLC) has drawn considerable attention recently from both the academy and industry [1, 2, 3]. VLC system usually consists of light emitting diode (LED) and photodiode (PD) as the optical front-end receivers. Indeed, due to the rapid development of LED technology, it has been possible to modulate the electrical information by LED in high frequencies not detectable by the human eye. Data transmission through VLC is achieved by the means of a simple and low-cost intensity modulation and direct/detection (IM/DD) [4], dealing with real and positive signals. Among the conventional modulation schemes investigated in the context of VLC are on-off-key (OOK), pulse-amplitude modulation (PAM) and multi-level pulse position modulation (M-PPM). Anyway, the performance of VLC systems is highly affected by different types of noise, deriving from external optical sources (e.g. artificial lights and sun) and noise derived from the hardware components, the filtering stages, etc. Even if this aspect represents one of the main factors delaying an extensive diffusion of VLC, relatively few authors in current literature have focused on the implementation of adaptive algorithms taking into account real time environmental condition evolution [5, 6].

In the system considered in this work, we consider three main sources of noise, thermal noise, amplifier noise and shot noise. The first two sources are independent of the signal and can be modeled as Gaussian distributions as shown in [7]. The shot noise can also be modeled by Gaussian distribution, except that it depends on the signal itself. Even though noise components highly impact on the quality of the received signal, few research works are reported in literature addressing the undesirable effects of the noise in VLC systems.

In [8] the authors propose an anti-noise modem for visible light communication systems. Their approach is based on M-ary position phase shift keying (MPPSK) modulation technique. The input MMPSK are analyzed and the interferences are then reduced by considering a set of improved waveform samples. In [9] a novel receiver as an ambient light rejection circuit is proposed. In particular, the authors implement an average-voltage tracking circuit at the receiver stage, in order to detect the average voltage produced by noise. The effects of noise on a VLC system are also described in [10]. In particular, authors show the pulse width distortions effects of the noise in messages coded by Manchester and Miller code. In [11] authors propose a variation of the compressed sensing method with partially aware support the reconstruct the clipping noise for the asymmetrically clipped optical OFDM systems.

In this paper, we propose an effective approach based on noise listening at the receiver stage. The received signal is processed both in the absence of data signal and in it presence. Hence, noise is then mitigated The transmitter stage is implemented by considering a software defined approach. The effectiveness of the software defined concept for visible light communication has been already assessed in [12, 13, 14, 15, 16, 17, 18]. Motivated by the outcome of these works, we have implemented different modulation techniques at the transmitting stage equipped with an USRP-N200.

The main contributions of this paper can be summarized as follows:

- We have proposed a signal processing noise mitigation approach taking into account its statistical features by analyzing also the effect of noise acquisition length;

- We implemented the approach on a real testbed by applying a software-defined approach based on the use of USRP platform;

The remainder of the paper is organized as follows. In the Section II we describe the system model and the hardware components for realizing our testbed based on USRP. In Section III we detail the noise mitigation approach. In Section IV we evaluate the performance of our approach. Finally, we conclude the paper in Section V.

## II. SYSTEM MODEL

In this Section we detail the transmitted signal and the specific components of our testbed.

In this work, we use Pulse Position Modulation for encoding transmitting bit stream. A single pulse is transmitted in one of the possible $M=2^b$ time slots which compound a symbol frame of duration T, where b is the number of bits carried by each frame. Considering this definition, the bit rate is $(1/MT)\log_2(M)$ [bit/s]. We tested our system using M=4 (4PPM) and M=8 (8PPM), while frame duration is T=1ms in both cases.

In each frame, the transmitted signal driving the LED can be expressed as a rectangular pulse of time length $q$, with time shift given by multiple integer of $\Delta$ given by

$$x(t) = A\Pi\left(\frac{t-(l+\frac{1}{2})\Delta}{q}\right) l \in [0, M-1]$$

If a fixed frame length T is chosen for each symbol representation, each pulse has a time length of $q=T/M$ and $q=\Delta$. In our case, pulses associate to each symbol have a time length of 0.250ms for 4PPM and 0.125ms for 8PPM. In both transmitting and receiving stage, signal is sampled with a sample rate equal to 1Msamples/s.

Dealing with the testbed, a schematic architecture of the system is outlined in Figure 1.

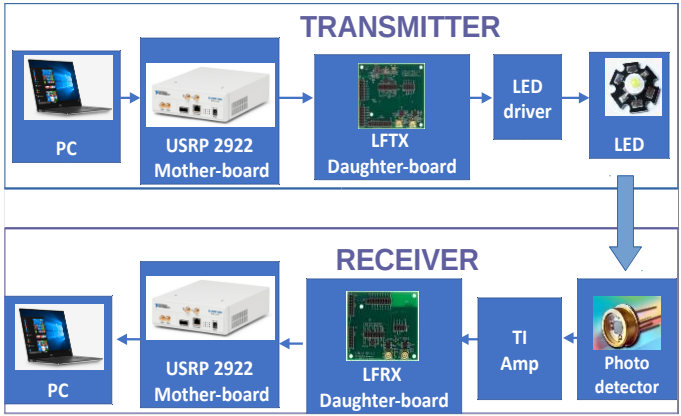

Figure 1. Architecture of the VLC communication system.

Two NI USRP 2922 devices have been employed in order to perform software defined operations. Each device is connected to a PC through a Gigabit Ethernet Interface. The commercial software LabView has been employed to manage signal processing, communication, measurements and graphical user interface. Two daughter-boards, a LFRX in the receiving VLC path and a LFTX in the transmitting stage, have been integrated for allowing the USRP 2922 to manage baseband signals. They also provide a filtering and signal conditioning stage. Clock generation and synchronization, ADCs and DACs are performed independently by the two motherboards. TX daughterboard is connected through an SMA cable to the driving circuit, which is composed of a Bias Tee (made up by a combination of inductances and capacitances), a voltage amplifier and 4 JFET transistors.

A low cost transmitting hardware front-end has been designed using an array of 4 commercial LED's, fed up by a 12 V DC power supply.

The receiving stage is composed of a commercial photodiode, a feeding and pre-amplifying network, allowing to lower DC component, let the diode to work in the correct current-voltage range and a trans-impedance amplifier (TIA), which converts photodiode output current to a voltage in the range [-1V,1V]. An SMA cable connects the output of the TIA to the receiving daughterboard. More details about the LED and the PD used in the experimental tests are provided in section IV. Since in Software defined paradigm all signal operations are performed by software, it is possible to employ several different modulations without any hardware modifications.

## III. EXTERNAL NOISE MITIGATION

The received signal $y(t)$ is processed in the electrical domain so it is properly sampled in order to allow noise processing and reduction. In detail, the sampled version of $y(t)$ is $y[n]$ and it is the overlap of different signals, like for external (optical) interference and thermal noise. While the thermal noise is generally white, thus meaning that samples are uncorrelated, the optical noise has its own *structure*, hence we can explore the autocorrelation properties so as to apply filtering, that is, prediction in order to mitigate its effect by subtraction.

Analytically speaking, we consider a first phase during which the interference is acquired. This requires that no transmission is active, and we have that the received signal is given by

$$y[n] = i[n] + w[n]$$

where $i[n]$ is the optical interference at the receiver represented in the electrical domain once converted by PD and $w[n]$ is the additive white Gaussian noise. We are interested in acquiring interference statistics with a special focus on autocorrelation that, once considered the received signal $y[n]$, it is given by

$$R_i[m] = R_y[n] - R_w[n]$$

that is, the autocorrelation of interference is the difference between the autocorrelation of received signal and noise. It is worth noting that since the noise is spectrally white (in a statistical sense), it is possible to measure it by neglecting interference (for example by posing an obstacle in front of the photodiode) through its power since it presents only a sample at m=0, whose amplitude is exactly the noise power level. From this latter, we can obtain the statistical features of the interference simply by evaluating the autocorrelation of the received sequence $R_y[m]$. As it will appear clearer in the following, having the opportunity to observe the interference for sufficiently long time can help the interference prediction and mitigation procedure. Hence, once we obtain the autocorrelation of the interference, we can proceed to predict it according to the following linear relationship

$$\tilde{\imath}[n] = \sum_{k=1}^{p} a_k i[n-k]$$

where $p$ is the order of linear predictor and the $a_k$ coefficients can be obtained according to the Yule-Walker equations [19] so as to have

$$\sum_{k=1}^{p} a_k R_i[j-k] = R_i[j]$$

Once the interference has been predicted, then is subtracted by the received signal so has to obtain a cleaner version of the information signal.

When transmission takes place by the transmitter, the (sampled) signal received is

$$r[n] = x[n] * h[n] + i[n] + w[n] - \tilde{\imath}[n]$$

where * denotes convolution between $x[n]$ and the channel $h[n]$. The detection utilizes a decision metric. Once the PPM masks are applied to the received signal, we obtain the decision metric

$$c_l = \boldsymbol{r}^T \boldsymbol{m}_l$$

where $\boldsymbol{r}^T$ is the column vector representing the signal received in a frame. while $\boldsymbol{m}_l$ is the row vector collecting the sample of the PPM mask matched on the $l$-symbol given in time by

$$m_l[n] = \Pi\left[n - (l + \frac{1}{2})\Delta\right]$$

$\Pi[n]$ being the pulse and $\Delta$ the time PPM symbol time shift above indicated. The Maximum-Likelihood decision rule is the described by the following relationship

$$\hat{l} = \arg \max_{l=0,\ldots,M-1} c_l$$

hence we select the symbol whose metric is the highest among all the possible M.

## IV. TEST RESULTS

We test our noise mitigation mechanism according to the framework shown in Figure 1.

TABLE I. TABLE 1 PHOTODIODE MODEL

| Photodiode Model | CENTRONIC OSD15-5T |
|---|---|
| Active Area | 15mm² |
| Responsivity (436nm) | 0.18-0.21 |
| Dark Current | 1-5 nA |

TABLE II. TABLE 2 LED MODEL

| Led Model | CREE MKRAWT-02-0000-0D00J2051 |
|---|---|
| Maximum Power | 15W |
| Maximum Light Flux | 1040 lm |
| Viewing Angle | 120° |

Photodiode features have been reported in Table I and Led characteristics in Table II.

The area of the detector is 15.5m x 9m and its exposed to outdoor light, on three sides of its perimeter (approximately oriented to Nord, South and West) as shown in Figure 2.

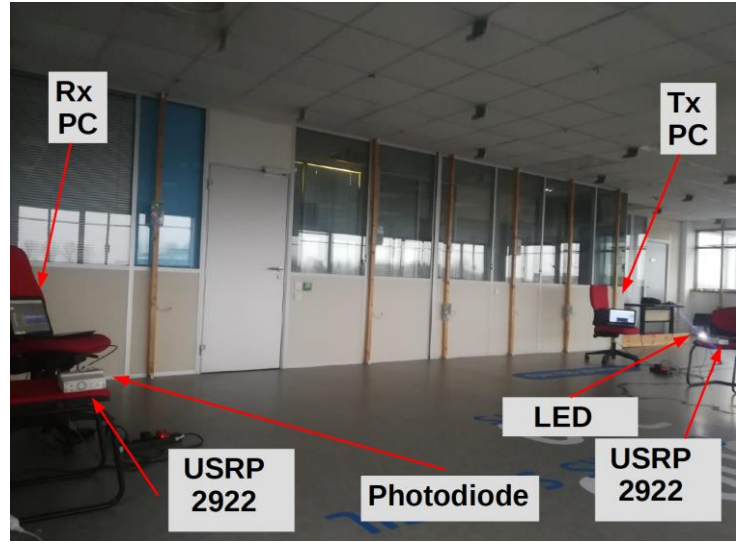

Figure 2: Scenario where the evaluation has been carried out.

Luminance produced by the LED array and sun light noise sources have been measured through an open source luxometer (Lux Light Meter Free), on a Huawei P8 Light smartphone (active screen area 11.5cm x 6.5cm) and then converted in lumen. These values have been reported, according to the distance between transmitter and receiver, in Table III.

TABLE III. DISTANCE (TX-RX) AND LUMINANCE

| Distance [m] | VLC LED Array Light Source (Luminance - lumen) |
|---|---|
| 2 | 296 |
| 4 | 200 |
| 8 | 27 |

We report in Figure 3, the sampled version of the received signal when no filtering is applied. It is possible to note in the upper part of the picture that the 4PPM signal is affected by interference due to external ambient lighting. This effect is considerably reduced in the lower part of Figure 3. Even though some residual interference is still present it is possible to recognize the typical structure of PPM signal. Moreover, it is remarkable that the signal dynamics, in the filtered case, considerably reduces. This is due to the amount of energy associated to the interference.

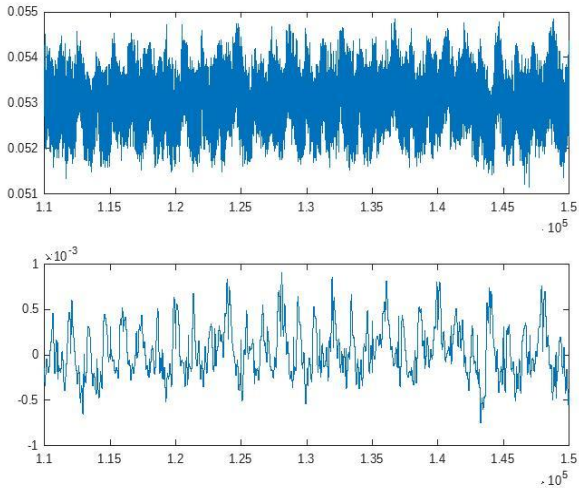

Figure 2. Example (upper figure) of the received signal in the presence of the external noise for 4PPM, at 4m and 100lux interference and effect of noise mitigation (lower figure).

Moving now to consider the effect of noise mitigation with respect to detection, we can observe in Figure 3 that symbol error rate (SER) increases when the interference strength increases. However, if we focus on solutions that do not consider filtering the interference, SER increases quickly when disturbing light increases its intensity. Higher is the distance between transmitter and receiver (ranging from 2mt to 8mt) higher is SER due to lower optical signal at the PD induced by channel attenuation.

A similar behavior is presented in the filtering approach described above. However, the gain offered with respect to the no filtering detection appears evident. The gain goes from one order of magnitude for an interference of 250lumen when 8mt is considered as distance between transmitter and receiver, till to a factor equating 30 when 2mt is considered as distance. It is worth noting that for interference lower than 150lumen, the filtering performance at 8mt are the same of 2mt with interference without its mitigation.

We report in Figure 4 the same kind of performance of those reported in Figure 3 with the only one difference of the adopted modulation format. In fact, there, 8PPM is implemented. The hierarchy of the performance of no filtering and filtering solutions, with respect to interference, is exactly the same of 4PPM. Interestingly, the order of magnitude of the gain offered by the filtering solution is essentially the same even though the values achieved by 8PPM with respect to SER are lower if compared with 4PPM.

The estimation of the autocorrelation and prediction of the interference is a crucial point since if autocorrelation acquisition is inaccurate, all the prediction procedure falls short and interference mitigation does not perform as expected. In order to emphasize this aspect, we report in Table I the SER obtained at 2mt when the statistics acquisition is performed with a different number of acquired samples when 4PPM is taken into account and 200lumen interference is considered.

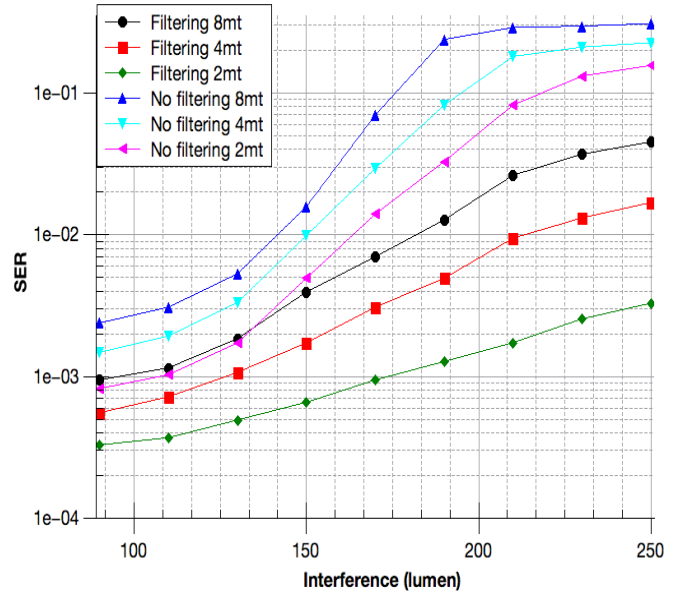

Figure 3. Performance of 4PPM modulation at different distances and with and without filtering

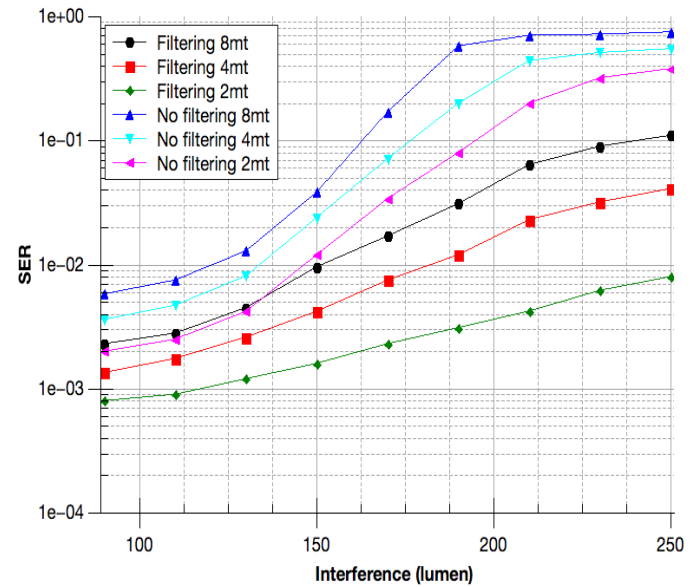

Figure 4. Performance of 8PPM modulation at different distances and with and without filtering

It is possible to infer from Table IV that while there is a strong different between capturing interference for only 10-250 samples, since SER is higher than 0.008, when we consider 2000 samples, SER falls down to 0.004. Even though 250 samples may appear a long delay for acquisition, at 1Ms/s sampling rate, it equates 250μs.

TABLE IV. SER AS A FUNCTION OF NUMBER OF SAMPLES COLLECTED FOR STATICAL ESTIMATION

| No of samples | SER |
|---|---|
| N=10 | 0.48 |
| N=50 | 0.32 |
| N=100 | 0.11 |
| N=250 | 0.008 |
| N=500 | 0.006 |
| N=1000 | 0.004 |
| N=2000 | 0.0039 |
| N=3000 | 0.0036 |
| N=4000 | 0.0036 |

## V. CONCLUSIONS

In this paper we have considered a Software-Defined VLC system, implemented through an USRP platform. One of the main issues of VLC system is represented by the different sources of noise interfering with the transmitted signal and impacting on the performance of the received signal. In order to limit the effect of the noise and in consideration of the existing of different source of disturbing signal, we have elaborated a simple while effective method for analyzing the noise and limit its effect on the signal. In particular, we "listen" the channel for a while in order to acquire sufficient information about the noise waveform. By treating the received signal with a signal processing method, we are able to reduce the impact of the noise and the results show that our method allows an high gain in different scenarios, at different distances between the transmitter and receiver and is available for different modulation techniques. In the specific context, we have validated the proposed approach by considering 4-PPM and 8-PPM, but the method is general and completely independent of the specific modulation technique considered.

## REFERENCES

[1] P.H.Pathak, X.Feng, P.Hu and P.Mohapatra, "Visible light communication, networking and sensing: A survey, potential and challenges", IEEE communications surveys and tutorials.vol.17.no.4.pp.2047-2077,2015

[2] C.H.Yeh, C.H.Chow, H.Y.Chen, Y.L.Liu and D.Z.Hsu, "Investigation of phosphor-LED lamp for real time half-duplex wireless VLC system", Journal of optics, vol.18, no.6,pp.1- 6,2016.

[3] D.Karunatilaka, F.Zafar, V.Kalavally, R.Parthiban, "LED based indoor visible light communications: state of the art", IEEE communications surveys and tutorials.vol.17.no.3.pp.1649-1677, 2015.

[4] Daniel J.F. Barros, Sarah K. Wilson, Joseph M. Kahn Comparison of orthogonal frequency-division multiplexing and pulse-amplitude modulation in indoor optical wireless links IEEE Trans Commun, 60 (1) (2012), pp. 153-163

[5] A. Cailean, B. Cagneau, L. Chassagne, M. Dimian, and Popa V. 2015. Novel Receiver Sensor for Visible Light Communications in Automotive Applications. IEEE Sensors Journal Optics Communications 15, 8 (2015), 44632–4639.

[6] C.W. Chow, C.H. Yeh, Y.F. Liu, P.Y. Huang, and Y. Liu. 2013. Adaptive scheme for maintaining the performance of the in-home white-LED. Optics Communications 292 (2013), 49–52.

[7] A. Lapidoth, S. M. Moser, and M. A. Wigger. 2009. On the capacity of free-space optical intensity channels. IEEE Transactions on Information Theory 55, 10 (2009), 4449–4461. https://doi.org/10.1109/TIT.2009.2027522

[8] Pu Miao, Lenan Wu, Zhimin Chen, "An anti-noise modem for visible light communication systems using the improved M-ary position phase shift keing, " in AEU – International Journal of Electronics and Communications, vol. 85, pp. 126-133, 2018.

[9] Quan Ngoc Pham, Vega Pradana Rachin, Jinyoung An and Wan-Young Chung, "Ambient Light Rejection Using a Novel Average Voltage Tracking in Visible Light Communication System," in Appl. Sci., n. 7, vol. 670, 2017.

[10] Alin Cailean, Barthélemy Cagneau, Luc Chassagne, Valentin Popa, Mihai Dimian. Evaluation of the noise effects on visible light communications using Manchester and Miller coding. International Conference on Development and Application Systems (DAS), Suceava, Romania, May 2014.

[11] Junnan Gao, Fang Yang, Sicong Liu and Jian Song, "Clipping noise cancellation based on compressed sensing for visible light communication," in Proceedings of the 3rd Workshop on Visible Light Communication Systems (VLCS'16).

[12] Q.Wang D.Giustiniano and D.Puccinelli,"OpenVLC:Software-defined Visible Light Embedded Networks," in *ACM VLCS*, 2014, pp. 15–20.

[13] E. Knightly, Y. Qiao, and H. Haas, "A software-defined visible light communications system with WARP," in Proc. 1st ACM Workshop Visible Light Commun. Syst., 2014, pp. 2434–2442.

[14] Q. Wang, D. Giustiniano, and D. Puccinelli, "OpenVLC: Software defined visible light embedded networks," in Proc. 1st ACM MobiCom Workshop VLCS, 2014, pp. 15–20.

[15] M. Rahaim, A. Miravakili, T. Borogovac, T. Little, and V. Joyner," Demonstration of a software defined visible light communication system," in Proc. 17th Annu. Int. Conf. Mobicom, 2011, pp. 1–4.

[16] Y.-S. Kuo, P. Pannuto, and P. Dutta, "System architecture directions for a software-defined lighting infrastructure," in Proc. 1st ACM MobiCom Workshop VLCS, 2014, pp. 3–8.

[17] A. Costanzo and V. Loscri, "A Learning Approach for Robust Carrier Recovery in Heavily Noisy Visible Light Communications," in Proceedings of IEEE Wireless Communications and Networking Conference, WCNC 2019.

[18] A. Costanzo and V. Loscri, "Visible Light Indoor Positioning in a Noise-aware Environment," in Proceedings of IEEE Wireless Communications and Networking Conference, WCNC 2019.

[19] Burg, J. P. (1968). "a new analysis technique for time series data". In *Modern Spectrum Analysis* (Edited by D. G. Childers), NATO Advanced Study Institute of Signal Processing with emphasis on Underwater Acoustics. IEEE Press, New York.